\documentclass[a4paper,journal]{IEEEtran}
\usepackage{geometry}
\usepackage{epsfig,latexsym,amssymb,amsmath,graphics,color}
\usepackage{graphicx,subfigure}
\usepackage[linesnumbered,ruled]{algorithm2e}
\usepackage{float}
\newtheorem{prop}{Proposition}

\newtheorem{cor}{Corollary}

\newtheorem{lm}{Lemma}

\newtheorem{thm}{Theorem}

\newcommand{\be}{\begin{eqnarray}}
\newcommand{\ee}{\end{eqnarray}}
\newcommand{\benn}{\begin{eqnarray*}}
\newcommand{\eenn}{\end{eqnarray*}}
\def\IR{\rm I \kern-0.20em R}

\newcommand{\bthm}{\begin{thm}}
\newcommand{\ethm}{\end{thm}}

\newcommand{\bcor}{\begin{cor}}
\newcommand{\ecor}{\end{cor}}
\newcommand{\bprop}{\begin{prop}}
\newcommand{\eprop}{\end{prop}}
\newcommand{\blm}{\begin{lm}}
\newcommand{\elm}{\end{lm}}
\newcommand{\beq}{\begin{equation}}
\newcommand{\eeq}{\end{equation}}
\newcommand{\ber}{\begin{eqnarray}}
\newcommand{\eer}{\end{eqnarray}}

\newcommand{\bproof}{\begin{proof}}
\newcommand{\eproof}{\end{proof}}

\newcommand{\bit}{\begin{itemize}}
\newcommand{\eit}{\end{itemize}}
\newcommand{\ben}{\begin{enumerate}}
\newcommand{\een}{\end{enumerate}}
\newcommand{\bdesc}{\begin{description}}
\newcommand{\edesc}{\end{description}}
\newcommand{\beqarrn}{\begin{eqnarray*}}
\newcommand{\eeqarrn}{\end{eqnarray*}}
\newcommand{\bproofof}{\begin{proofof}}
\newcommand{\eproofof}{\end{proofof}}
\newenvironment{rem}{\begin{trivlist}\item[]{\bf
Remark:}\hspace{4mm}}{\end{trivlist}}
\newcommand{\brem}{\begin{rem}}
\newcommand{\erem}{\end{rem}}
\newenvironment{rems}{\begin{trivlist}\item[]{\bf
Remarks}\begin{itemize}}{\end{itemize}\end{trivlist}}
\newcommand{\brems}{\begin{rems}}
\newcommand{\erems}{\end{rems}}
\newtheorem{fact}{Fact}
\newcommand{\bfact}{\begin{fact}}
\newcommand{\efact}{\end{fact}}
\newtheorem{examp}{Example}
\newcommand{\bexamp}{\begin{examp}\rm}
\newcommand{\eexamp}{\end{examp}}
\newtheorem{defn}{Definition}
\newcommand{\bdefn}{\begin{defn}\rm}
\newcommand{\edefn}{\end{defn}}

\newtheorem{alg}{Algorithm}
\newcommand{\balg}{\begin{alg}}
\newcommand{\ealg}{\end{alg}}

\newtheorem{prob}{Problem}
\newcommand{\bprob}{\begin{prob}}
\newcommand{\eprob}{\end{prob}}

\newcommand{\bvtm}{\begin{verbatim}}
\newcommand{\bfig}{\begin{figure}}
\newcommand{\efig}{\end{figure}}
\newcommand{\bcen}{\begin{center}}
\newcommand{\ecen}{\end{center}}

\long\def\comment#1{}

\def \n2{{N_0 \over 2}}

\def \h5{\hspace{0.5in}}

\def\IR{\mathbb R}

\renewcommand{\baselinestretch}{1.0}
\renewcommand{\arraystretch}{1.0}

\renewcommand{\baselinestretch}{1}

\title{Pulse-laser Based Long-range Non-line-of-sight Ultraviolet Communication with Pulse Response Position Estimation}

\author{Ruixiong Xu, Chen Gong, and Zhengyuan Xu
}

\date{}
\begin{document}
\maketitle{}

\renewcommand{\baselinestretch}{1.0}

\begin{abstract}
 We propose pulse laser-based ultra-violet communication over long distance, such that the pulse response signals can be detected at the receiver at the cost of low data transmission rate. We characterize the signal and achievable performance for the pulse laser-based communication. Since the detection performance critically depends on the pulse response position estimation, we also propose two approaches to estimate the pulse response positions, one based on counting the number pulses in a window, and the other based on the correlation of pulse response shape and the number of detected photoelectrons. It is seen that the correlation-based position estimation approach can achieve more accurate estimation compared with the counting-based one.
\end{abstract}


\section{Introductions} \label{sec.Introduction}
Ultraviolet (UV) optical wireless communication~\cite{OptCommun} serves as a candidate for the mobile scenario since it does not require perfect alignment of the transmitter and receiver~\cite{ExtRangeNLOS,MagzineNLOS} and the UV background radiation is negligible. UV communication can maintain a certain data rate in a certain local area, which can also satisfy certain secrecy requirements. Most non-line of sight (NLOS) channel models adopt stochastic geometry framework and Monte Carlo simulation to evaluate the link gain~\cite{ding2009modeling,xu2008analytical,wang2010non}, and the path loss of NLOS channel has been studied with single scattering approximation~\cite{ding2009modeling}. In the past few years, extensive experimental works~\cite{ModelingNLOSSingle,liao2015uv} have been conducted on the NLOS channel models, mostly on short ranges. Long-range UV scattering channels have been characterized in representative work~\cite{liao2015long}.

In the perspective of communication, experimental communication systems with transmission data rate of 2kbps over 100m distance have been reported in~\cite{siegel2004short}. Realtime systems with approximately 2.4kbps air-face data rate over 90m transmission links, experiment with 500kbps air-face data rate over 50m NLOS transmission links based on RS/LDPC codes and semi-real-time system with nearly 80kbps air-face data rate over 30m NLOS transmission links have been accomplished in~\cite{han2012theoretical,wu2014experimental,guo2015experimental}, respectively. Based on simulation results with measured parameters, the link budget of nearly 1Mbps throughput over 80m NLOS transmission is performed in~\cite{qin2017received}. Furthermore, real-time demonstration of 400kbps throughput over 500m transmission link based on concatenated convolutional and RS codes has been realized in~\cite{wang2017demonstration}. Real-time communication over $1$km NLOS transmission link with $1$Mbps throughput have been reported in~\cite{wang20181mbps}. Based on the measurement, it was predicted that the data rate could reach 100kbps at a range of 2000 meters~\cite{liao2015uv,liao2015long}.

In this work, we consider even longer transmission distance. We adopt pulse laser for communication, where certain output response signal can still be observed under a large path loss. The cost of the long transmission distance is the low data rate, since the rate of a pulse laser is typically within one hundred pulses per second. We analyze the achievable performance on the pulse laser based UV communication. As the receiver-side processing involves the localization of the pulse response, we also propose two approaches for the pulse response position estimation, based on counting the number of pulses and the correlation of pulse response shape and the number of detected photoelectrons. It is seen that the latter one shows more accurate pulse position estimation if the pulse response shape is known.

\section{Signal and System Model} \label{sec.System}

\subsection{Ultraviolet Communication Channel Characterization}
According to the theory of Mie scattering and Rayleigh scattering, a multiple scattering Monte Carlo (MC) model is constructed, which can be further simplified using single scattering approximation.
In such a model, the new propagation direction can be depicted by two angles, the scattering angle $\theta_s$ and the azimuth angle $\phi$. Note that the scattering angle reflects the distributions of photons emitted in different directions for a large number of photons passing through the scattering point. The phase functions of Mie scattering and Rayleigh scattering, respectively, are given as follows,
\be \label{equ.System1}
P^M(\mu) =\frac{{1-g^2}}{4\pi}[\frac{{1}}{({1+g^2-2g\mu})^{3/2}}+f\frac{{3\mu^2-1}}{2({1+g^2})^{3/2}}],
\ee
\be \label{equ.System2}
P^R(\mu) =\frac{3[{1+3\gamma+(1-\gamma){\mu^2}}]}{16\pi({1+2\gamma})},
\ee
where $\gamma$ and $f$ are model parameters in the ray-tracing simulation; $g$ represents the mean of $\cos\theta_s$; and $\mu=\cos\theta_s$ is defined from the scattering angle $\theta_s$.

We adopt weighted sum of Rayleigh scattering and Mie scattering phase functions as the scattering phase function, given by
\be \label{equ.System3}
P(\mu) =\frac{{k^{R}_s}}{k_s}P^R(\mu)+\frac{{k^{M}_s}}{k_s}P^M(\mu),
\ee
where $k^R_s$ and $k^M_s$  are Rayleigh scattering coefficients and Mie scattering coefficients, respectively; and the total scattering coefficient $k_s=k^R_s + k^M_s$.

We adopt random free distance $r_0$ to characterize the photon propagation between two consecutive scattering events. For a single photon, the traveling distance to the next scattering point obeys an exponential probability density function with mean $1/{k_e}$. Then, we have
\be \label{equ.System4}
r_0 =-\frac{\ln\xi_s}{k_s},
\ee
where $k_s$ is the extinction coefficient of the atmosphere as the sum of scattering coefficient and absorption coefficient, and random variable $\xi_s$ satisfies the uniform distribution between 0 and 1. Note that $\theta_s$ is obtained by solving the following equation,
\be \label{equ.System5}
\xi_\mu =2\pi\int_{-1}^\mu P(\mu)d\mu ,
\ee
where $\mu=cos\theta_s$; $\xi_\mu$ is uniformly distributed between 0 and 1; and $ P(\mu)$ is given by Equation (\ref{equ.System3}). The probability of each photon successfully arriving at the receiver is calculated by integrating the phase function with the scattering angle, depending on the elevation angle and receiver FOV.
\renewcommand\arraystretch{0.9}
\begin{table}
\centering
\caption{Parameters in Simulation.}\label{tab.Table1}
\begin{tabular}{|p{5cm}|p{3cm}|}
  \hline
  Parameter & Value \\
  \hline
  Wavelength $\lambda$ & 266nm \\
  \hline
  Absorption  coefficients  $k_a$ & $0.74\times10^{-3}{m^{-1}}$ \\
  \hline
  Rayleigh scattering coefficients $k^{R}_s$ & $0.2456\times10^{-3}{m^{-1}}$ \\
  \hline
  Mie scattering coefficients $k^{M}_s$ & $0.25\times10^{-3}{m^{-1}}$ \\
  \hline
  Receiving aperture area & $1.77\times10^{-4}{m^{2}}$ \\
  \hline
  Mie phase function asymmetry parameter g& 0.72 \\
  \hline
  Angular field of view of receiver & $\pi/6$ \\
  \hline
  Elevation angle of the transmitter & $\pi/3$   \\
  \hline
  Elevation angle of the receiver & $\pi/3$  \\
  \hline
\end{tabular}
\end{table}

\subsection{Pulse Broadening for Long-Distance UV Communication}
The pulse broadening when propagating through the NLOS UV scattering channel enables the pulse laser-based long-distance NLOS UV communication. Here we investigate the relationship between receiver elevation angle and pulse broadening. As each measured impulse response exhibits random variations, we average sufficiently many impulse responses per geometry to obtain the average response. The pulse broadening is calculated based on left and right boundaries on the impulse response of UV signals.

Key system parameters on the pulse broadening simulation are shown in Table~\ref{tab.Table1}. Figure~\ref{fig.Fig1} shows the pulse broadening for different receiver elevation angles (from 0-$\pi/2$) for receiver distance $5$Km, where each time the elevation angle is increased by $\pi/12$, It can be seen that the pulse broadening is not sensitive to the elevation angle.

\begin{figure}
  \centering
  \includegraphics[width=9.5cm, height=7cm]{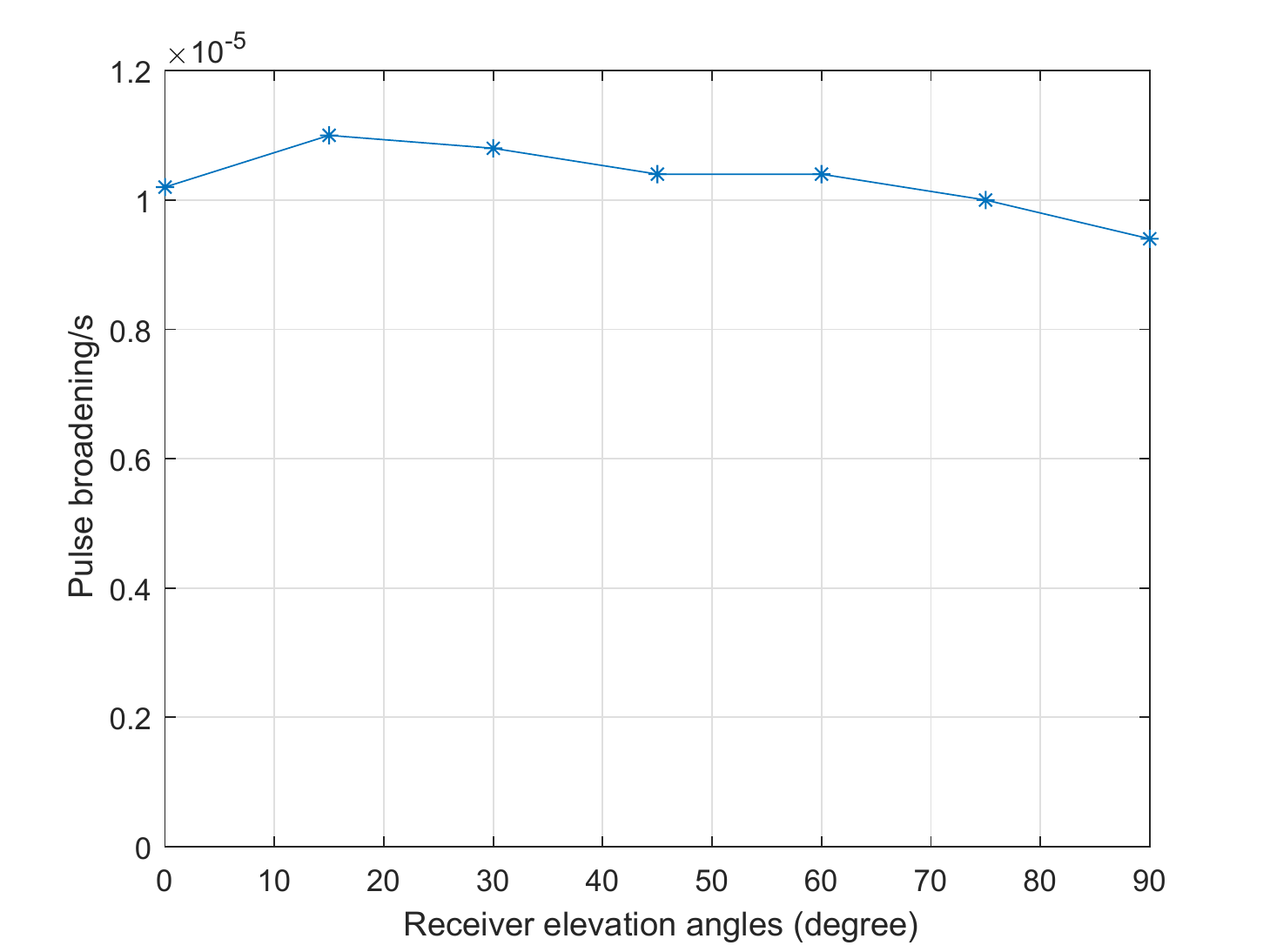}\\
  \caption{Pulse broadening for different receiver elevation angles.}\label{fig.Fig1}
\end{figure}

\subsection{Pulse-based Communication System Block} \label{sec.DetForward}

\begin{figure}
  \centering
  \includegraphics[width=9.5cm, height=4.2cm]{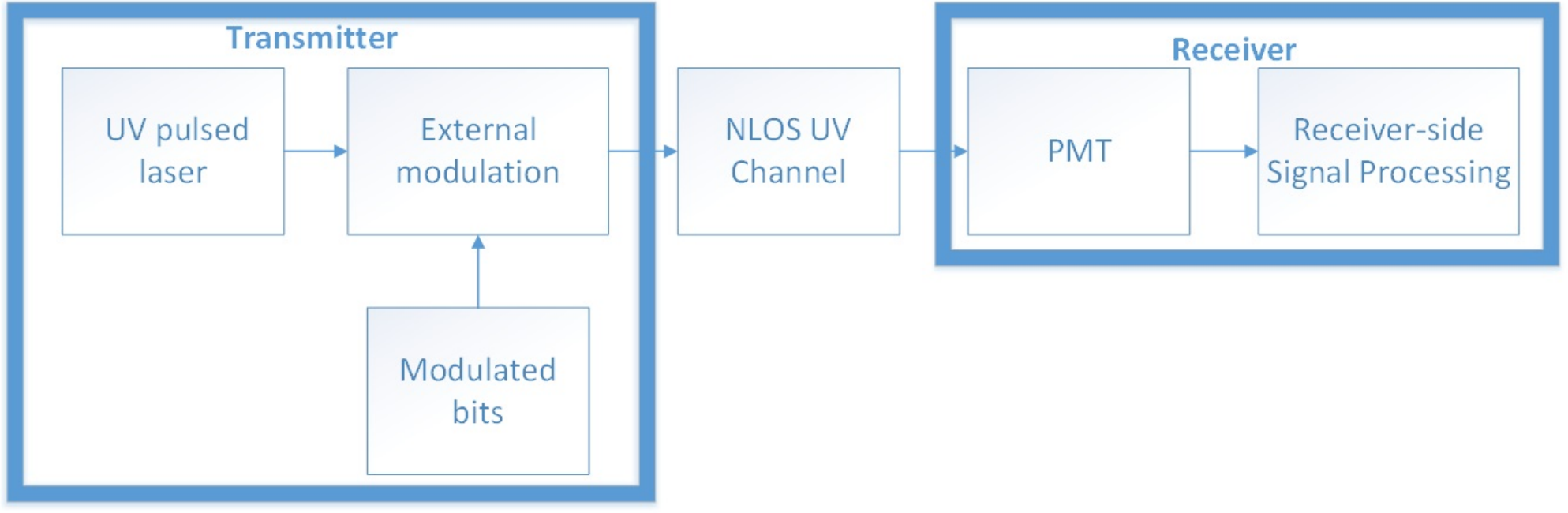}\\
  \caption{Block diagram for the pulse-based UV communication system.}\label{fig.Fig2}
\end{figure}

We consider the pulse-based optical wireless scattering communication, whose system block diagram is shown in Figure~\ref{fig.Fig2}. At the transmitter side, a pulse laser is adopted to generate periodical laser pulses. The coded bits are fed into an external modulator controlling the on-off state of the external modulator. The pulse can be transmitted to the receiver if and only if the external modulator is on. Since the pulse rate is typically low, a physical mechanics based external modulator can be employed, for example a metal pipe with a mechanical on-off switch. The signals are detected by a PMT and then processed by digital signal processing techniques.

The main motivation on the pulse-based communication lies in large path loss due to long transmission distance. If a continuous laser/LED transmitter is adopted, the received signal energy will spread in a long window with larger background radiations, which degrades the communication system performance. If using a pulse laser transmitter, the received signal energy can concentrate into a shorter window, with lower background radiation photons that can improve the communication performance, provided that the window position can be accurately estimated.

\section{Received Signal Characterization and Link Budget Evaluation} \label{sec.Characterization}
\subsection{Received Signal Characterization}
Due to the strong attenuation of wireless optical scattering channel, the received optical signal can be characterized by discrete photoelectrons. The intensity of photon arrival at a certain time is proportional to the amplitude of pulse response, which varies over time. The background radiation can be treated as a constant-rate Poisson arrival process with arrival rate $\Lambda_b$. Assume that the energy per pulse is given by $E$, and $E_p=h\nu$ is the energy per photon, where $h$ and $\nu$ denote Planck¡¯s constant and the optical signal frequency, respectively. Let $g(t)$ denote the pulse response from the transmitter to the receiver, in the form of received power, and $\eta$ denote the quantum efficiency of the detector. Then, the Poisson arrival rate of photoelectrons, denoted as $\Lambda_s(t)$, is given by
\be \label{equ.System7}
\Lambda_s(t)=\frac{\eta g(t)}{E_p}.
\ee
The mean number of photoelectrons within a pulse response, denoted as $\lambda_s$, is given by the integral of $\Lambda_s(t)$ within that pulse response interval.
\subsection{Achievable System Performance}
\begin{figure}
  \centering
  \includegraphics[width=9.5cm, height=7cm]{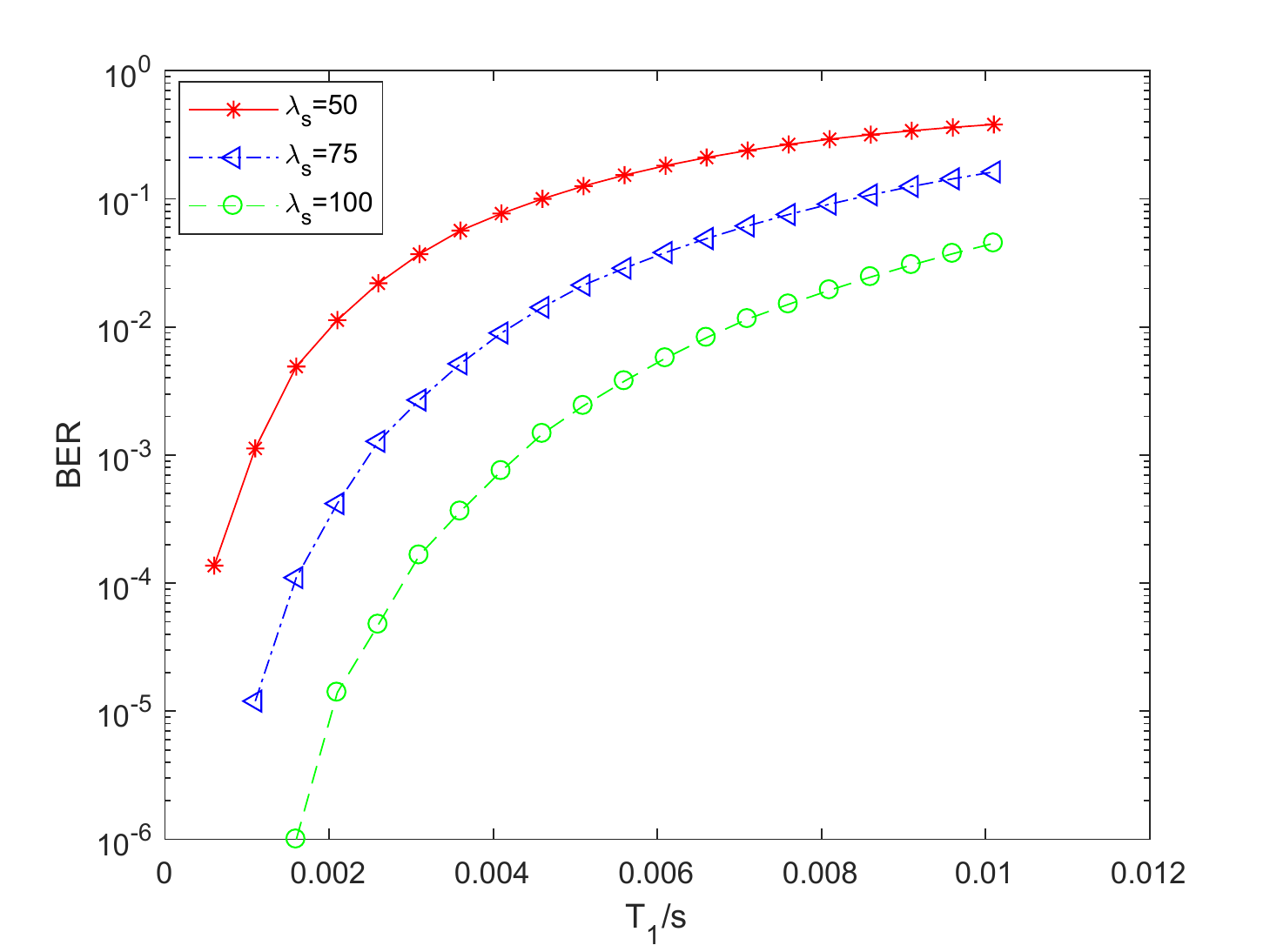}\\
  \caption{BER performance with respect to the pulse processing window length $T_1$.}\label{fig.Fig3}
\end{figure}

Assume that the length of processing window is $T_1$, where the pulse response is located in that window and the pulse response is performed via counting the number of photoelectrons in that window. The bit error rates (BERs) under different lengths $T_1$ and for different mean numbers of photoelectrons for the signal components $\lambda_s$ are shown in Figure~\ref{fig.Fig3}. According to the link gain simulation results, $\lambda_s$ ranges from $10^1 \thicksim 10^2$ photons for the transmission distance longer than $5$Km. Thus, we select the mean numbers of photoelectrons $\lambda_s=50,75,100$ and background radiation rate $\Lambda_b=5\times10^4$ photons per second. Note that the mean number of background radiation photons is given by $\lambda_b = \Lambda_bT_1$. It can be seen that the BER is below $10^{-3}$ for the signal component $\lambda_s > 50$ and processing time $T_1 < 1ms$. More sophisticated processing approaches based on the pulse response can be adopted to further reduce the BER.

One performance comparison benchmark is the UV communication based on continuous laser. Assuming symbol rate $R$ and energy per pulse $E$, the corresponding continuous laser-based communication has the transmission power $ER$ with processing window length $1/R$. Then, the pulse-laser based communication may outperform the continuous laser-based one if the processing time window length $T_1 < 1/R$, which can be readily achieved in real practical applications.

\section{Pulse Localization} \label{sec.ALGORITHM}
Note that accurate localization of pulse response under the discrete pulse-type signals can reduce the detection BER. We propose two pulse response localization approaches, one based on simply counting the number of photoelectrons in case of unknown pulse response at the receiver, and the other based on matching the number of photoelectrons with the pulse response, in the case of known pulse response shape at the receiver.

\subsection{Two Pulse Localization Approaches}

\subsubsection{Counting the Total Number of Pulses}
We set a sliding window of length $T$, and count the number of pulses in that window. Let $N$ denote the number of pulses in that sliding window. The start of a pulse is treated as the first time $N$ higher than a threshold, and the end of a pulse is treated as the first time $N$ lower than that threshold after the start of a window. The pulse counting threshold can be selected based on Neyman-Pearson criterion, under a certain false alarm probability since the mean number of photoelectrons $\lambda_b = \Lambda_bT$ is known. A typical value of the threshold is two, i.e., the start of a window is detected if and only if there are more than two photoelectrons in that window.

%

\subsubsection{Matching with the Pulse Response}
 In case of fixed distances and angles of the transmitter and receiver, the pulse response shapes through the same channel are similar. We can adopt such pulse shape to estimate the pulse response position. We consider the chip structure of length $\tau_c$. We perform a sliding correlation operation with the pulse shape sequence, which can be obtained via averaging over sufficiently many random channel realizations. The length of the pulse shape sequence is the longest of the generated realizations. Let $Z_i$ denote the number of photoelectrons detected in each chip and $S(m)$ denote the  pulse shape sequence. The sliding window correlation finds the position where the correlation exceeds a threshold and reaches the maximum as the start of a pulse response, and the first time it falls down to zero after the start of a window as the end of a pulse response. The correlation is computed as follows,
\be \label{equ.System9}
Correlation(i)=\sum_{m=1}^M Z_{i+m}\cdot{S(m)}.
\ee



\subsection{Numerical Results}
Assume that the  distance varies from $5$Km to $9$Km, and the same geometry configuration parameters as shown in Table I. The real starting time and ending time of a pulse response are denoted as $P_L$ and $P_R$, respectively.

\subsubsection{Counting the Total Number of Pulses}
We set the counting window length to $T=0.02\times 10^5$. The estimated starting time and ending time for one channel realization are denoted as $\hat P_{L1}$ and $\hat P_{R1}$, respectively, as shown in Table~\ref{tab.Table2}.
\renewcommand\arraystretch{0.9}
\begin{table*}
\centering
\caption{True and estimated Time Window under different distance parameters (Unit Second).}\label{tab.Table2}
\begin{tabular}{|p{4cm}|p{2cm}|p{2cm}|p{2cm}|p{2cm}|p{2cm}|}
  \hline
  distance/km & 5.0  & 6.0 & 7.0 & 8.0 & 9.0 \\
  \hline
  $P_L$ & $8.000\times10^{-6}$ & $8.000\times10^{-6}$ & $8.000\times10^{-6}$ & $8.000\times10^{-6}$& $8.000\times10^{-6}$ \\
  \hline
  $\hat P_{L1}$ & $8.000\times10^{-6}$ & $8.060\times10^{-6}$ & $8.340\times10^{-6}$ & $8.060\times10^{-6}$ & $8.000\times10^{-6}$ \\
  \hline
  $\hat P_{L2}$ & $8.060\times10^{-6}$ & $7.980\times10^{-6}$ & $7.980\times10^{-6}$ & $8.120\times10^{-6}$ & $7.980\times10^{-6}$ \\
  \hline
  $P_R$ & $1.242\times10^{-5}$ & $1.374\times10^{-5}$ & $1.304\times10^{-5}$  & $1.208\times10^{-5}$ & $1.202\times10^{-5}$ \\
  \hline
  $\hat P_{R1}$ & $1.212\times10^{-5}$ & $1.372\times10^{-5}$ & $1.234\times10^{-5}$ & $1.208\times10^{-5}$ & $1.146\times10^{-5}$  \\
  \hline
  $\hat P_{R2}$ & $1.242\times10^{-5}$ & $1.372\times10^{-5}$ & $1.302\times10^{-5}$ & $1.208\times10^{-5}$ & $1.202\times10^{-5}$  \\
  \hline
  $e_M$ for counting-based & $1.238\times10^{-6}$  & $1.241\times10^{-6}$ & $1.196\times10^{-6}$ &$1.338\times10^{-6}$ &$1.056\times10^{-6}$ \\
  \hline
  $e_M$ for matching-based & $0.521\times10^{-6}$  & $0.713\times10^{-6}$ & $0.558\times10^{-6}$ &$0.338\times10^{-6}$ &$0.714\times10^{-6}$ \\
  \hline
\end{tabular}
\end{table*}

\subsubsection{Matching with the Pulse Response}


\begin{figure}
  \centering
  \includegraphics[width=9cm, height=7cm]{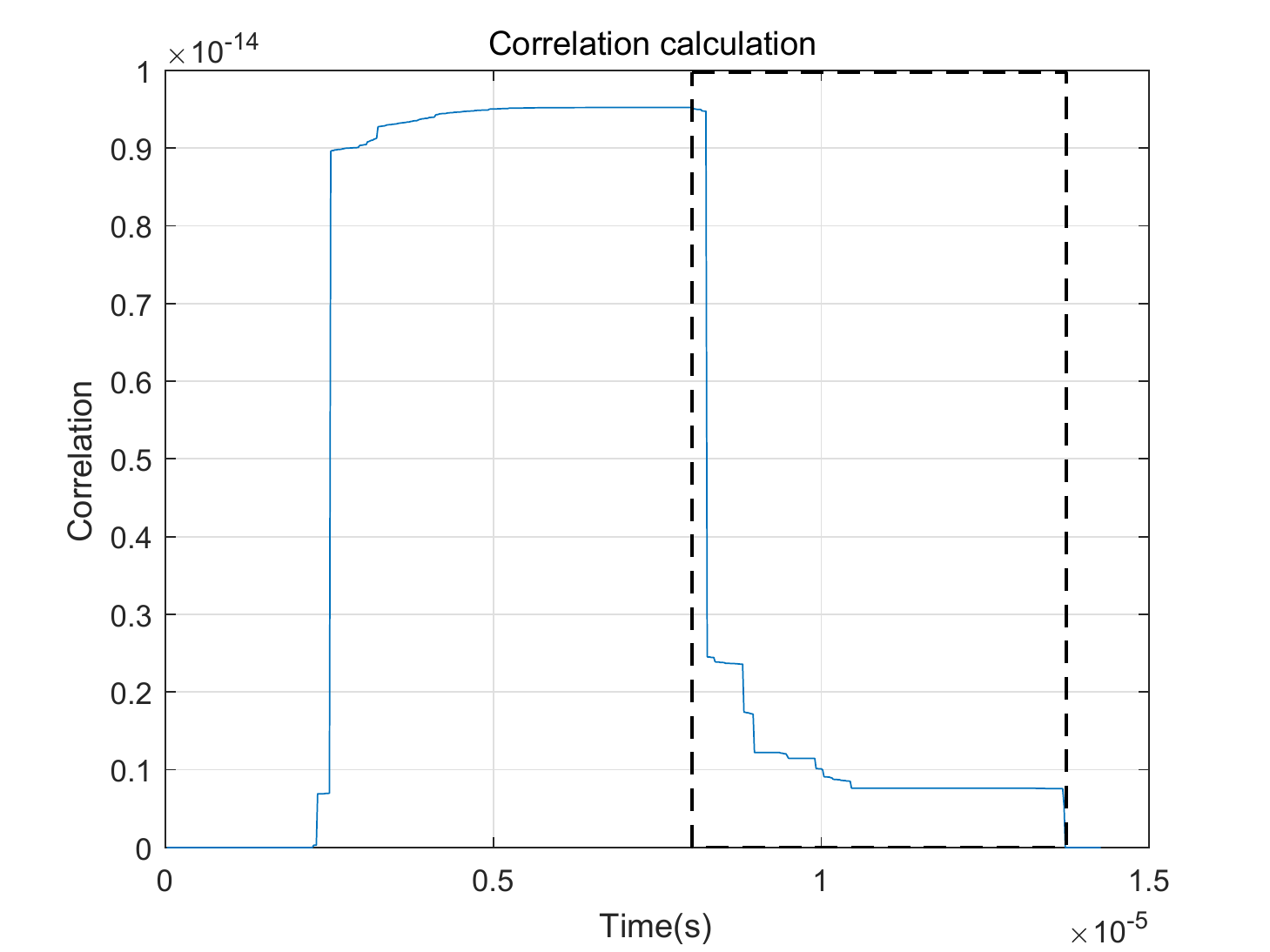}\\
  \caption{The value of correlation using the sliding window.}\label{fig.Fig6}
\end{figure}

The estimated starting time and ending time for one channel realization are denoted as $\hat P_{L2}$ and $\hat P_{R2}$, respectively, as shown in Table~\ref{tab.Table2}. Figure~\ref{fig.Fig6} shows the values of correlation in a dotted line, for one channel realization under transmission distance $6$Km. When the correlation reaches the maximum, the left pulse response position estimate is given by $\hat P_{L2}=7.980\times10^{-6}$; and when the correlation reaches zero, the right pulse response position is given by $\hat P_{R2}=1.372\times10^{-5}$. Considering the real pulse response positions $(P_{L2},P_{R2})=(8.000\times10^{-6},1.374\times10^{-5}) $, the matching-based position estimate is more accurate compared with the counting-based one.

We use the mean absolute deviation to characterize the accuracy of the pulse position estimation, denoted as $e_M$. The mean of absolute deviation for the starting time and ending time estimation over $1000$ channel realizations are shown in Table II. It is seen that the correlation-based pulse position estimation shows lower deviation compared with the pulse-counting based one.

\section{Conclusions} \label{sec.Conclusions}
We have proposed pulse laser-based UV communication over long distance. We have investigated the signal characterization and achievable communication performance. We have also proposed two approaches on the pulse response position estimation, where the correlation-based approach can achieve more accurate estimation compared with the pulse counting-based one.
\renewcommand{\baselinestretch}{1.0}
\bibliographystyle{./IEEEtran}
\bibliography{./gong_18}

\end{document}